\documentclass{article}

\usepackage{arxiv}

\usepackage[utf8]{inputenc} 
\usepackage[T1]{fontenc}    
\usepackage{hyperref}       
\usepackage{url}            
\usepackage{booktabs}       
\usepackage{amsfonts}       
\usepackage{nicefrac}       
\usepackage{lipsum}		
\usepackage{graphicx}
\usepackage[numbers]{natbib}
\usepackage{doi}
\usepackage{authblk} 
\usepackage{amsmath} 

\usepackage{comment}

\title{Gaussian copula modeling of extreme cold and weak-wind events over Europe conditioned on winter weather regimes}

\author[1, 2, *]{Paulina Tedesco}
\author[3]{Alex Lenkoski}
\author[4, 5]{Hannah C. Bloomfield}
\author[6, 7]{Jana Sillmann}

\affil[1]{Department of Physics, University of Oslo, Oslo, Norway}
\affil[2]{Information Technology Department, Norwegian Meteorological Institute, Oslo, Norway}

\affil[3]{Statistical Analysis and Machine Learning, Norwegian Computing Center, Oslo, Norway}

\affil[4]{School of Geographical Sciences, University of Bristol, Bristol, UK}

\affil[5]{Department of Meteorology, University of Reading, Reading, UK}

\affil[6]{Center for Earth System Research and Sustainability, Research Unit for Sustainability and Climate Risks, Universität Hamburg, Hamburg, Germany}

\affil[7]{CICERO Center for International Climate Research, Oslo, Norway} 

\affil[*]{Corresponding author: Paulina Tedesco, \textit {paulina.tedesco@gmail.com}}

\renewcommand{\shorttitle}{Gaussian copula modeling of cold and weak-wind events}

\hypersetup{
pdftitle={A template for the arxiv style},
pdfsubject={q-bio.NC, q-bio.QM},
pdfauthor={David S.~Hippocampus, Elias D.~Striatum},
pdfkeywords={First keyword, Second keyword, More},
}

\begin{document}
\maketitle

\begin{abstract}
A transition to renewable energy is needed to mitigate climate change. In Europe, this transition has been led by wind energy, which is one of the fastest growing energy sources. However, energy demand and production are sensitive to meteorological conditions and atmospheric variability at multiple time scales. To accomplish the required balance between these two variables, critical conditions of high demand and low wind energy supply must be considered in the design of energy systems. We describe a methodology for modeling joint distributions of meteorological variables without making any assumptions about their marginal distributions. In this context, Gaussian copulas are used to model the correlated nature of cold and weak-wind events. The marginal distributions are modeled with logistic regressions defining two sets of binary variables as predictors: four large-scale weather regimes and the months of the extended winter season. By applying this framework to ERA5 data, we can compute the joint probabilities of co-occurrence of cold and weak-wind events on a high-resolution grid (0.25 deg). Our results show that a) weather regimes must be considered when modeling cold and weak-wind events, b) it is essential to account for the correlations between these events when modeling their joint distribution, c) we need to analyze each month separately, and d) the highest estimated number of days with compound events are associated with the negative phase of the North Atlantic Oscillation (3 days on average over Finland, Ireland, and Lithuania in January, and France and Luxembourg in February) and the Scandinavian Blocking pattern (3 days on average over Ireland in January and Denmark in February). This information could be relevant for application in sub-seasonal to seasonal forecasts of such events.
\end{abstract}



\section{Introduction}
Affordable and clean energy is one of the UN Sustainable Development Goals (SDGs). Energy is also crucial for achieving many of the other SDGs. The energy sector currently accounts for more than two-thirds of the global greenhouse gas emissions \cite{irena_2020}. Consequently, a decarbonization of the energy sector is required to meet the SDGs and the Paris agreement \cite{gwec_2017}, by increasing the share of renewable power generation \cite{Rogelj_et_al}. Europe is leading this transition, although it is still one of the world's biggest energy consumers and greenhouse gas emitters \cite{liobikine_butkus}. 

The production of clean energy is highly weather-dependent. Very cold or warm temperatures increase the demand due to heating and cooling, respectively \cite{taylor_buizza_2003}. Extreme weather conditions, such as European blocking in wintertime, can lead to high electricity demand and low renewable power production (energy shortfall). How to deal with periods of low renewable production is a big challenge in the design of renewable energy systems \cite{huber_et_al_2014}. Better subseasonal and seasonal forecasts help improve decision-making and planning \cite[e.g.,][]{orlov_et_al_2020}. They are of value for power producers to better prepare for extreme meteorological events, and could be an important planning tool for traders, plant operators, and investors for managing climate variability related risk \cite{cortesi_et_al_2019}, as they provide relevant information for price forecasting \cite{pinson_2013}. Although studies have traditionally focused on single drivers \cite{zscheischler_et_al_2018}, more recent research is considering the interactions between them \cite[e.g.,][]{van_der_Wiel_2019, bloomfield_etal_2020, drucke_et_al_2021, kaspar_et_al_2019, turner_et_al_2019, otero_et_al_2021}.

 Given that electricity production and demand depend on the weather, they vary on multiple timescales \cite[e.g.][]{sinden_2007, besseq_fouquau_2008, Bloomfield_2016}. Hence, it is important to understand how large-scale circulation systems influence peaks of demand and energy production to identify periods of over- and under-supply. Over time, the impact of large-scale atmospheric patterns on extreme winter temperatures mostly does not change, despite slow fluctuations of these patterns \cite{schuhen_et_al_2022}. Meteorological conditions preceding energy shortfall are described as anomalous high pressure systems combined with below normal temperatures \cite{van_der_wiel_2019_b, Bloomfield_2018, van_der_Wiel_2019}. Moreover, moderate energy droughts (with duration and severity exceeding the 75th percentile), are expected every half a year \cite{otero_et_al_2021}. Further research is still needed to understand the impact of atmospheric variability on surface variables that combined lead to low supply. As far as we know, no other research has proposed a meteorological based methodology to model the bivariate probabilities of cold spells and weak-winds.

Several indices exist that describe European climate variability through daily synoptic-scale weather patterns. A method that has been proven to be useful in weather forecasting and climate change applications is the computation of weather regimes (WRs) with the k-means algorithm \cite[e.g.,][]{neal_et_al_2016, ferranti_et_al_2015, matsueda_palmer_2018, michelangeli, cassou_2008}. Understanding the surface impact of these WRs is relevant for subseasonal to seasonal energy applications \cite{van_der_Wiel_2019, bloomfield_etal_2020, bloomfield_et_al_2021, cassou_2010}. Furthermore, the response of power systems to these patterns across Europe has recently been studied, with a focus on the North Atlantic Oscillation (NAO) \cite[e.g.,][]{ely_et_al_2013, thornton_et_al_2017, Bloomfield_2018, brayshaw_et_al_2011, zubiate_et_al_2016}. 

Modeling the dependence between demand and energy production is of key importance to understand the occurrence of energy shortfall and prevent it by redesigning the energy systems. However, modeling multivariate distributions can be a challenge. Copulas offer a powerful and flexible tool that returns the joint probability of events as a function of the marginal probabilities of each event. This makes copulas attractive, as the univariate marginal behavior of random variables can be modeled separately from their dependence. The semi-parametric Bayesian Gaussian copula method used here estimates multivariate relationships between variables with univariate marginal distributions that cannot be well approximated with a simple parametric model \cite{hoff_2007}, making it a promising candidate for modeling the joint probabilities of cold and weak-wind events. This method has been used in the post-processing of a multivariate mesoscale weather forecast \cite{moller_et_al_2013}. 

Our goal is to develop a methodology for diagnosing the probabilities of meteorological compound events associated with high energy shortfall in the winter, when electric heating is highest \cite{van_der_wiel_2019_b}. In this study, we focus only on wind power and do not assess solar or hydro. We apply the methodology to cold and weak-wind events over Europe and analyze the results obtained over the continent, where local onshore wind power generation can meet a fraction of the demand. First, we show that the WRs improve the estimation of marginal probabilities of low temperature and weak wind events and that these events are correlated. Then, with Gaussian copulas, we model the dependency between cold and weak-wind events as a function of their marginals, separately from their dependence, and we show where the models are significantly improved when accounting for the correlations between the events. Furthermore, the strong spatial and intraseasonal variability indicates that we need to analyze every month separately on a grid level.

The paper proceeds as follows. Section \ref{data_and_methods} describes the data and the methodology developed to compute the probabilities cold and weak-wind events. Section \ref{results} presents the results of the multivariate probabilities, as well as the univariate probabilities and the associations between cold spells and weak-wind events. Finally, discussions and conclusions are provided in section \ref{discussion_and_conclusions}. Additional figures and their respective analysis can be found in the appendix.


\section{Data and Methods} \label{data_and_methods}

This section describes the data and methods used to estimate joint probabilities of cold and weak-wind events conditioned on the WRs using a Gaussian copula framework. 

\subsection{Reanalysis data}\label{data}

Temperature is well established as the main weather driver of electricity demand. We use hourly 2 m temperature and 10 m wind data from the ERA5 reanalysis \cite{hersbach_et_al_2019}, produced by the European Centre for Medium-Range Weather Forecasts (ECMWF). The data are provided on a regular latitude-longitude grid at $0.25^\circ \times 0.25^\circ$, and the period covered in this work is from 1979--2021. We compute daily minimum temperatures from these hourly data. Wind speeds are calculated from the eastward (u) and northward (v) components at an hourly frequency and then aggregated to daily maximum values. Daily minimum wind speeds are usually close to zero, but low daily maximum values represent well conditions for poor wind power generation. Despite the good coverage and long records, and the fact that ERA5 outperforms other global reanalyses frequently used in the literature, it is discouraged to use global reanalyses to estimate mean winds because of the uncertainty in these data \cite{ramon_et_al_2019}.

\subsection{Weather regimes}\label{weather_regimes}

Atmospheric circulation is well known for its variability at multiple time scales being reflected in weather patterns and circulation systems. These WRs are quasi-stationary large-scale circulation patterns \cite{reinhold_pierrehumbert}. They typically persist for 6--10 days, are spatially well defined, and limited in number. The understanding of the causes of their recurrence, persistence, and transition is crucial for medium-range, subseasonal, and seasonal-to-interannual climate prediction \cite{cassou_2008, cassou_2010}, as they influence the weather at the surface and, subsequently, the renewable power generation and electricity demand \cite{grams_et_al_2017, thornton_et_al_2017, bloomfield_etal_2020}.

WRs are obtained here by applying the k-means algorithm with four centroids on geopotential height at 500 hPa (Z500) data \cite{michelangeli, cassou_2008}. The algorithm leads to four winter regimes in the Euro-Atlantic area (27$^\circ$N--81$^\circ$N, 85.5$^\circ$W--45$^\circ$E), and has been adopted by other authors \cite[e.g., ][]{van_der_Wiel_2019, bloomfield_etal_2020}. The Z500 fields of the obtained regimes are similar, although not identical, to the teleconnection patterns defined by the National Oceanic and Atmospheric Administration (NOAA), and can be interpreted as the negative and positive phases of the NAO, the Atlantic Ridge (AR), and the Scandinavian Blocking (SCAND) (see figure \ref{fig:WRs} in appendix \ref{sec:appendix_composites}). We will therefore use these names to refer to the four WRs. Moreover, it has been shown that temporal sub-sampling \cite{cassou_2008} and the use of different reanalysis data \cite{bloomfield_etal_2020, van_der_Wiel_2019} do not change the spatial structure of the regimes nor the optimal partition ($k = 4$).

The classification method consists of two steps. First, a cosine weight as a function of latitude is applied to the Z500 data and the first fourteen Empirical Orthogonal Functions (EOFs) are computed. The associated principal components (PCs) were used as coordinates of a reduced phase space. Then, the PCs are clustered into four groups with the K-means algorithm, choosing 30 random starts and a maximum of 100 iterations. Every daily map is assigned to a centroid based on its closest Euclidean distance. The number of regimes, $k=4$, corresponds to the most robust regime partition during winter months \cite{michelangeli}.

\subsection{Logistic regression}
In this study, we use logistic regressions to classify cold and weak-wind events. The predictors consist of two sets of indicator variables, one for the five winter months and one for the four WRs. The variables in each group are dichotomous and take only the values 1 or 0. The univariate distributions will later be used in the copula framework to estimate the joint probabilities. 

Logistic regressions are widely used linear models for binary classification. Let $Y$ be the binary outcome variable indicating failure or success with ${0, 1}$. Then, $p$ stands for the probability of a positive event, i.e., $p = P(Y=1)$. The mathematical expression of the logit function is:

\begin{equation}
logit(p) = log(\frac{p}{1-p}).
\end{equation}

It is assumed that the logit transformation of the outcome variable has a linear relationship with the predictor variables: $x_1, x_2, ..., x_k$. Then $\beta_0, \beta_1, ..., \beta_k $ are the parameters estimated via the maximum likelihood method when performing a logistic regression of Y on $x_1, x_2, ..., x_k$:

\begin{equation}\label{eq:logisticReg}
logit(p) = log(\frac{p}{1-p}) = \beta_0 + \beta_1 x_1 +  ... + \beta_k x_k.
\end{equation}

We are usually interested in predicting the probability that a particular sample belongs to a particular class. The formula of the probability $P(Y=1)$ is:

\begin{equation}\label{eq:sigmoidFunc}
p = \frac{e^{\beta_0 + \beta_1 x_1 + ... + \beta_k x_k}}{1+e^{\beta_0 + \beta_1 x_1 + ..., \beta_k x_k}} = \frac{1}{1+e^{-(\beta_0 + \beta_1 x_1 + ... + \beta_k x_k})},
\end{equation}

\subsection{Copulas}
Our goal is to establish a methodology for modeling the probabilities of co-occurrence of cold temperatures and weak winds. However, estimating joint densities is not an easy task since only a few non-Gaussian families are defined, and non-parametric estimation is demanding. Nonetheless, density estimation in one variable is relatively easy, given that many convenient families exist (e.g., logistic, exponential, uniform) and that the non-parametric approach is efficient and accurate. The copulas framework for modeling multivariate distributions provides a flexible representation and separates the univariates from their true nature of dependence.

In the field of probability theory and statistics, a copula function $C:[0, 1]^n \rightarrow [0, 1]$ is defined as a multivariate distribution

\begin{equation}\label{eq:copula_fun}
    C(u_1,u_2,....,u_n) = P(U_1 \leq u_1, U_2 \leq u_2, ... U_n \leq u_n), 
\end{equation}

such that marginalizing gives $U_i \sim Uniform(0,1)$. Copulas are useful because we can transform any arbitrary random variable into a uniform and back. The function that transforms uniforms to any other univariate distribution is the inverse of the cumulative distribution function (CDF). To do the opposite transformation, from an arbitrary distribution to the uniform(0, 1), we apply the inverse of the inverse CDF, the CDF.

\subsubsection{Gaussian copulas} \label{gaussian_copulas}

Copula theory ensures that, for every joint multivariate distribution, there exists a unique copula. In the case of the Gaussian copula function, finding its parameters is limited to finding the correlation matrix of the random variables we want to study.

A Gaussian copula is given by 

\begin{equation}\label{eq:gaussian_copula}
   C(u_1, u_2, ..., u_n) = \Phi_\Sigma(\Phi^{-1}(u_1), \Phi^{-1}(u_2), ..., \Phi^{-1}(u_n)),
\end{equation}

where $\Phi_\Sigma$ represents the CDF of a multivariate normal with covariance $\Sigma$ and mean $0$, and $\Phi^{-1}$ is the inverse CDF for the standard normal.

Given a multivariate distribution 

\begin{equation}
    F_X(X) = P(X_1 \leq x_1, X_2 \leq x_2, ..., X_n \leq x_n) = \Phi_\Sigma(x_1, x_2, ..., x_n),
\end{equation}

we can extract its Gaussian copula

\begin{equation}\label{eq:fx_distribution}
    \begin{split}
       F_X(X)  &= \Phi_\Sigma(F_1^{-1}(F_1(X)), F_2^{-1}(F_2(X)), ..., F_n^{-1}(F_n(X)))  \\
       & = \Phi_\Sigma(F_1^{-1}(u_1), F_2^{-1}(u_2), ..., F_n^{-1}(u_n)) \\ 
       & = \Phi_\Sigma(\Phi^{-1}(u_1), \Phi^{-1}(u_2), ..., \Phi^{-1}(u_n))  \\ 
       & = C(\Phi^{-1}(u_1), \Phi^{-1}(u_2), ..., \Phi^{-1}(u_n)) 
    \end{split}
\end{equation}

and plug in any marginal into the copula function.

The inverse CDF transforms the uniforms into normal distributions. Then, the multivariate normal's CDF squashes the distribution so that the marginals are uniform and with Gaussian correlations. Thus, the Gaussian Copula is a distribution over the unit hypercube $[0, 1]^n$ with uniform marginals.

\subsubsection{Semiparametric copula estimation}
It is often the case that the marginal distributions do not belong to standard families. In such cases, it might be appropriate to use a semi-parametric strategy that involves representing the associations among variables with a simple parametric approach and estimating the marginals nonparametrically. 

We use an extended rank likelihood method of semiparametric inference for copula, which is a function of the association parameters only \cite{hoff_2007}. It can be applied without any assumptions of the marginal distributions, making it appropriate for the joint analysis of continuous and ordinal discrete data. The R package 'sbgcop' \cite{sbgcop} provides a tool for estimation and inference for the Gaussian copula parameters.

\subsection{Model performance}\label{model_performance}
We use permutation tests to evaluate the significance of the results at a $10 \%$ level in terms of the BSS. In the logistic regression framework, the BSS is used to assess the performance of models against a seasonal baseline model that does not include any information about the WRs (the only predictors are the indicator variables for the months). The BSS is also used to evaluate the copula model's performance against an independent model with no correlation.

The most common measure for probabilistic forecasts is the Brier Score (BS) \cite{brier_score, wilks}. It is assumed that the events only can occur in one class on each of the $n$ occasions. For dichotomous events, the forecast probabilities are $f_{i}$ for an occasion $i$ and the BS is

\begin{equation}\label{eq:brier_score_bin}
BS = \frac{1}{n} \sum_{i=1}^{n} (f_{i} - E_{i})^2,
\end{equation}

where $E_{i}$, takes the value 1 if the event occurred and 0 otherwise. The score is negatively oriented, meaning that a perfect forecast has a $BS=0$. Less accurate forecasts exhibit higher scores, but since individual forecasts and observations are both bounded by zero and one, the range of possible values for the BS is $0 \leq BS \leq 1$. 

The BSS is often used and, since $BS_{perfect} = 0$, it takes the form

\begin{equation}\label{eq:brier_skill_score}
BSS = \frac{BS - BS_{ref}}{0 - BS_{ref}} = 1 - \frac{BS}{BS_{ref}}.
\end{equation}

Negative values mean that the forecast is less accurate than the reference forecast; when the forecast presents no skill compared to the reference $BSS = 0$; and a perfect skill compared to the reference forecast reflects in a skill score equal to 1.

\section{Results} \label{results}

We first compute the marginal probabilities and the correlations between cold and weak-wind events. Then, we analyze the joint probabilities, which can be interpreted as the percentage of days we expect compound events for a given month and regime.

\subsection{Marginal probabilities of cold and weak-wind events} \label{marginals}

Marginal probabilities of cold events and weak-wind events are modeled for each pair of regime and month using logistic regressions. 

\begin{figure}
    \centering
    \includegraphics[scale=0.35]{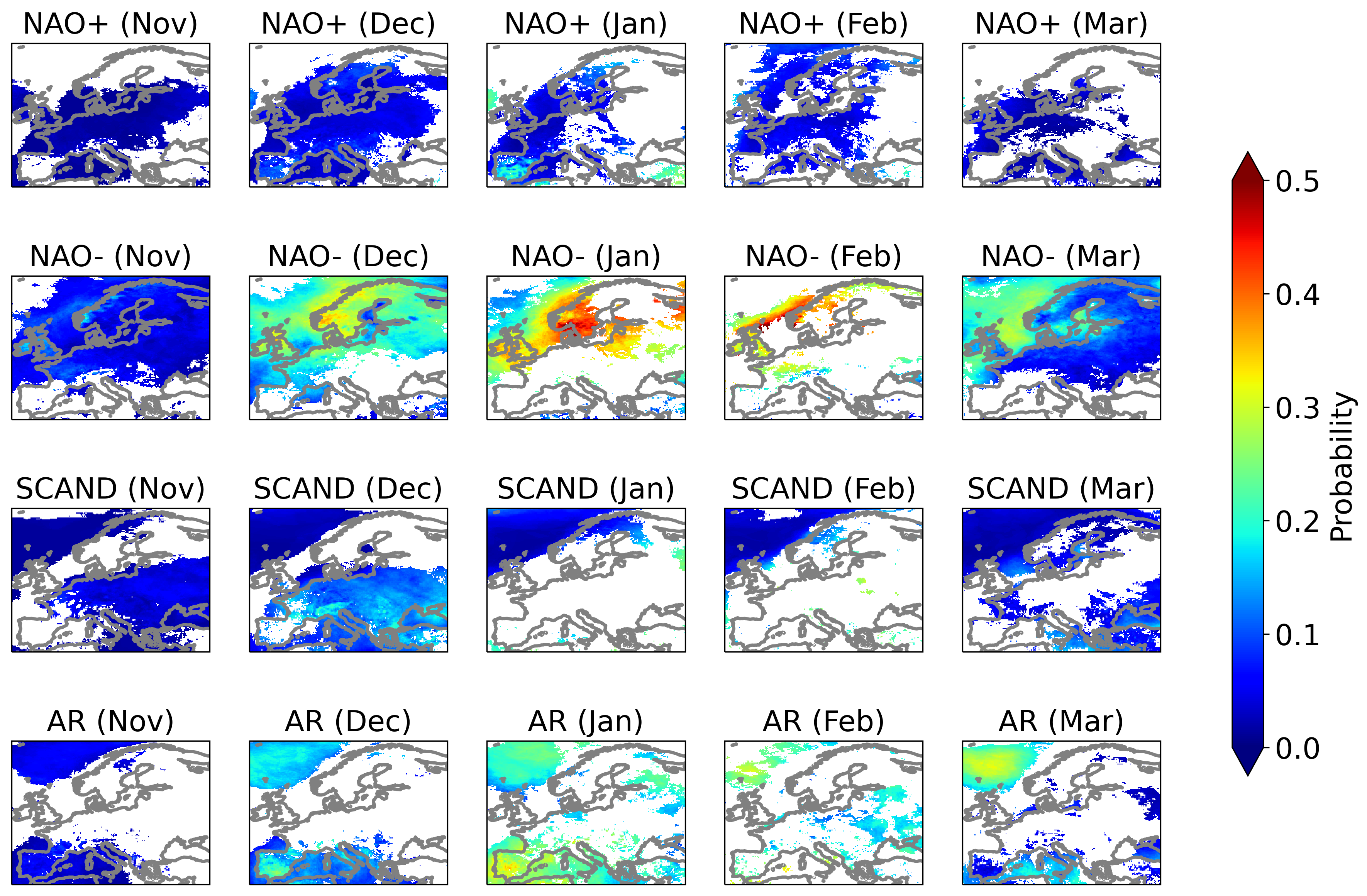}
    \caption{Marginal probabilities of cold events where the BSS is significant. Colors show the probabilities of cold events for each month and weather regime (WR) estimated with logistic regressions. Cold events are defined for each grid box as days with daily minimum temperature below $10^{th}$ percentile computed for the entire winter season (NDJFM). Probability values where the BSS is negative or not significant at a 10\% level are masked out. The BSS was computed with respect to a climatology model. Area of study: 35$^\circ$N--72$^\circ$N, 11$^\circ$W--40$^\circ$E. Figure based on ERA5 data (1979--2021).}
    \label{fig:logistic_t2m_masked}
\end{figure}

Figure \ref{fig:logistic_t2m_masked} illustrates the regions where the BSS computed for cold events is significant, i.e., where the WRs provide relevant information to the logistic model. It shows, in addition, the probabilities of observing temperatures lower than the $10^{th}$ in these regions. What stands out from figure \ref{fig:logistic_t2m_masked} is that regions with significant BSS cover extended areas. Even when probabilities are close to zero, like in November or over continental areas in March, considering the WRs improves the performance of the logistic regressions used to model cold events. If we focus on continental areas where the BSS is significant, we observe probability values higher than 0.3 over Scandinavia for NAO- and over the Iberian Peninsula for AR. Furthermore, although the areas with significant BSS are similar throughout the season for a given regime, the strong intraseasonal variability of the probabilities indicates that it is convenient to analyze the temperature events for each month separately rather than for the whole season. 

A strong negative NAO index is usually associated with cold winter temperatures across Northern Europe and Northern Asia, which agrees with the pattern observed for the NAO- cluster in figure \ref{fig:logistic_t2m_masked}. Provided that the $10^{th}$ percentile was computed at each grid box for the entire season (NDJFM), the highest probabilities are observed during the coldest months. Probabilities are higher than 0.2 from December to February over land, in smaller areas of Northern Europe and Scandinavia, and from December to March over the sea, which we speculate is due to a slower response of the water. 

The AR pattern is characterized by an anticyclonic circulation over the Atlantic with a weak surface temperature anomaly over Western Europe (see figure \ref{fig:composites_t} in appendix \ref{sec:appendix_composites}). Despite the weak anomaly pattern, the AR is the regime with the second highest significant probabilities of cold events, reaching local probability values higher than 0.3 across the Iberian Peninsula in January. Furthermore, the pattern shows a significant signal in the North Atlantic, with probabilities increasing from November to March. Still, we are mostly concerned about strong signals over land, which is the case of the Iberian Peninsula in December and January. 

The other two regimes, SCAND and NAO+, exhibit probabilities close to zero when the BSS is significant and positive. Thus, for NAO+ and SCAND, the model is only improved by the WRs in regions where cold events are less likely to occur.

\begin{figure}
    \centering
    \includegraphics[scale=0.35]{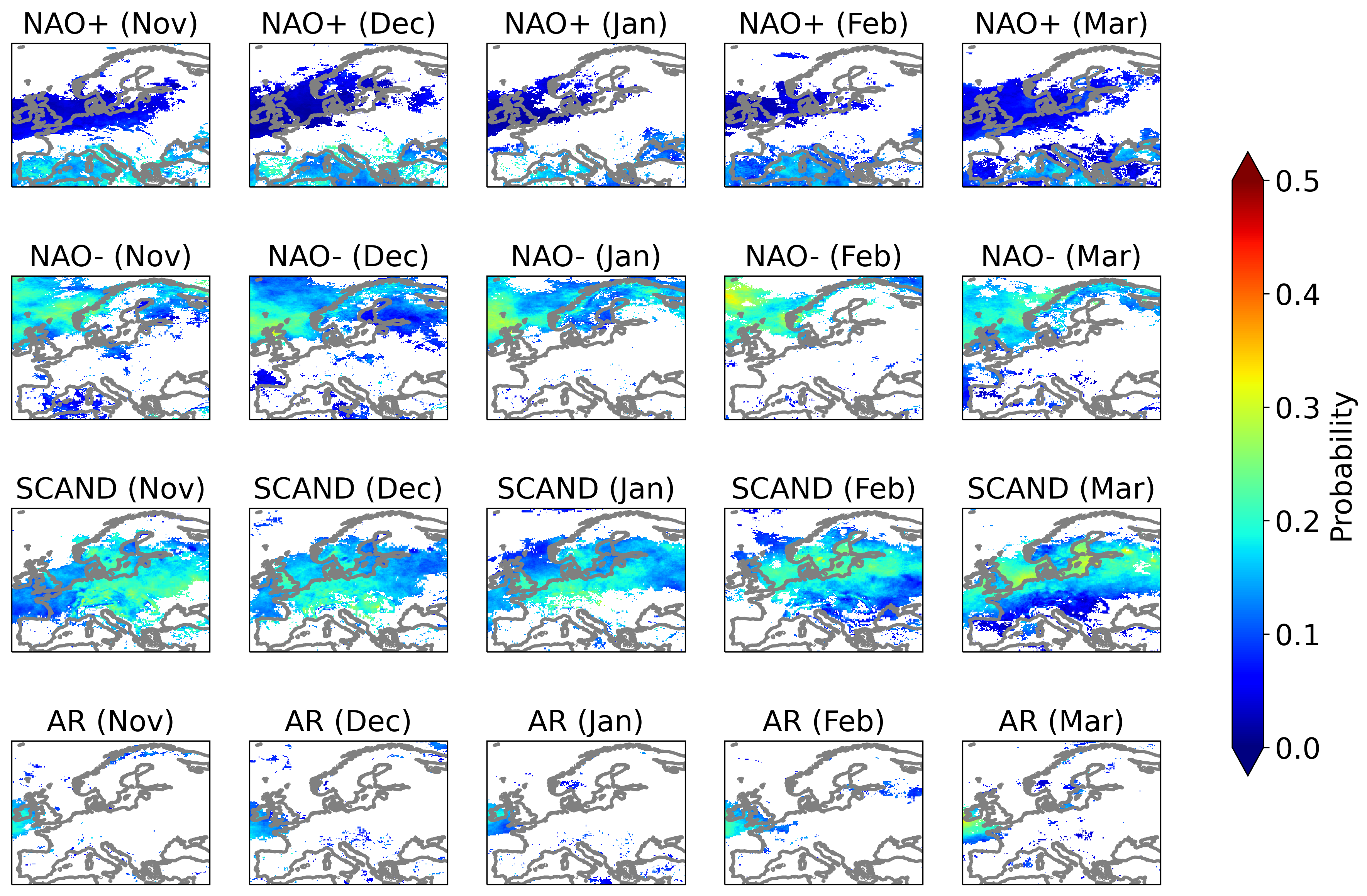}
    \caption{Marginal probabilities of weak-wind events where the BSS is significant. Colors show the probabilities of weak-wind events for each month and weather regime (WR) estimated with logistic regressions. Weak-wind events are defined for each grid box as days with daily maximum wind speeds below $10^{th}$ percentile computed for the entire winter season (NDJFM). Probability values where the BSS is negative or not significant at a 10\% level are masked out. The BSS was computed with respect to a climatology model. Area of study: 35$^\circ$N--72$^\circ$N, 11$^\circ$W--40$^\circ$E. Figure based on ERA5 data (1979--2021).}
    \label{fig:logistic_ws10m_masked}
\end{figure}

Figure \ref{fig:logistic_ws10m_masked} is analogous to figure \ref{fig:logistic_t2m_masked} but has been computed for daily maximum wind speeds lower than the $10^{th}$ percentile. This variable presents persistent zonal bands of significant probabilities at different latitudes when NAO+, NAO-, or SCAND are dominant. However, significant probabilities of weak-wind events are lower than 0.3 for all the regimes. The weather regimes improve the results mainly over the North Atlantic and Scandinavia during NAO-, and in Northern and Central Europe during SCAND. On the other hand, NAO+ shows two zonal bands of significant probabilities, one with probabilities under 0.2 over the Mediterranean and one with probabilities close to zero over Northern Europe.  

In general, we observe that the marginal probabilities obtained for the NAO-, NAO+, and SCAND patterns support earlier results \cite{bloomfield_etal_2020, van_der_Wiel_2019}. However, the monthly probabilities calculated for AR conditions indicate a stronger response of the temperatures to large-scale upper-level circulation than the seasonal aggregated results reported in previous studies \cite{bloomfield_etal_2020, van_der_Wiel_2019}. Given that November is the warmest month of the winter season and fewer temperature events are registered, the results for this month excluded from the analysis in the following. We have also excluded the results for NAO+ for the same reason. In addition, figures of unmasked marginal probabilities are provided in appendix \ref{sec:appendix_unmasked_probabilities} for both variables.

\subsection{Correlations between cold and weak-wind events}

The correlations between cold and weak-wind events are the parameters of the Gaussian copulas used to model their joint distributions. We compute these correlations with the 'sbgcop' R package \citep{sbgcop}.

\begin{figure}
    \centering
    \includegraphics[scale=0.35]{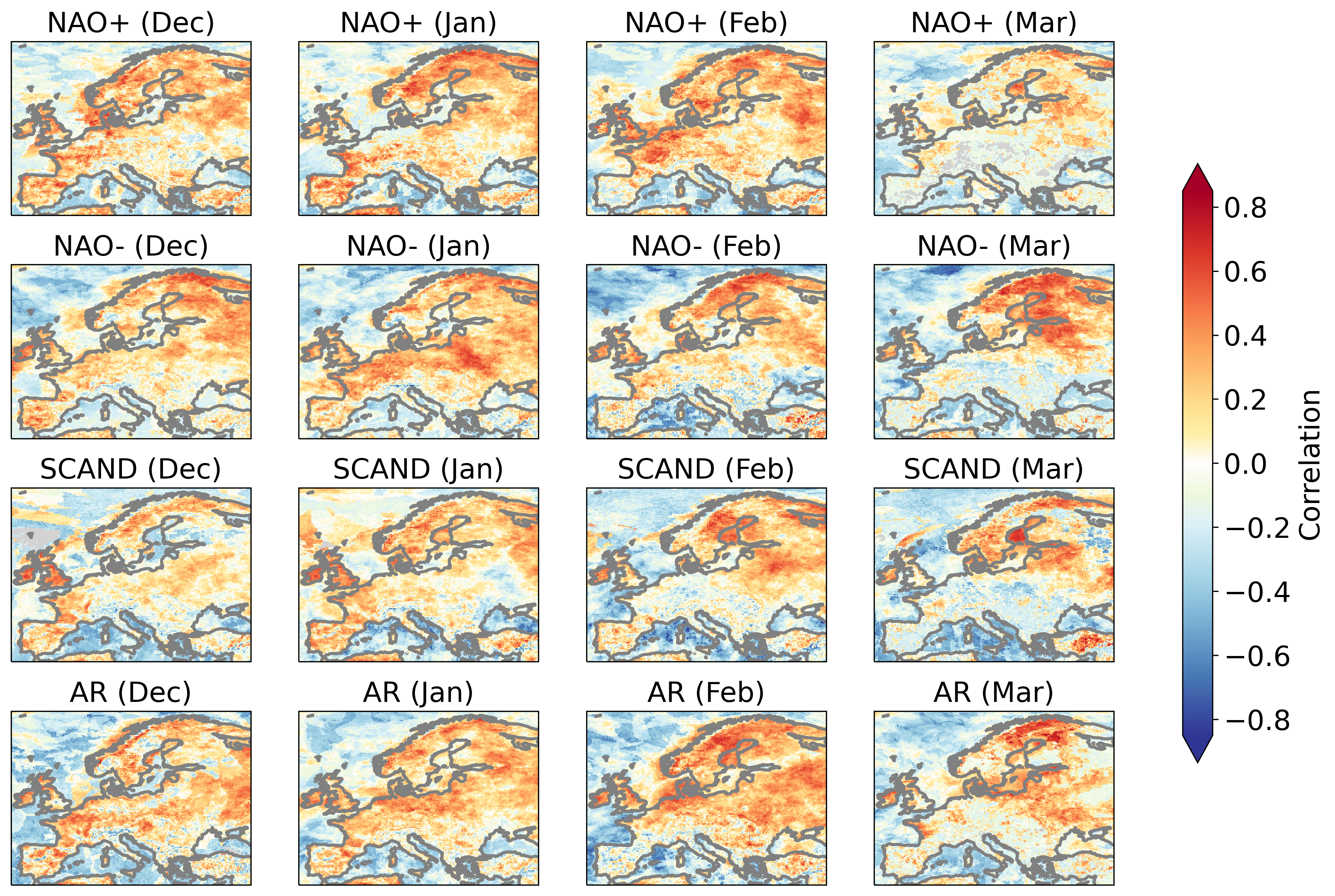}
    \caption{Correlations between cold events and weak-wind events for each Weather Regime (WR) and month. The threshold that defines the events at each grid box is the $10^{th}$ percentile over the whole season. Locations with no register of cold events are represented in gray. Area of study: 35$^\circ$N--72$^\circ$N, 11W--40E. Figure based on ERA5 data (DJFM, 1979--2021).}
    \label{fig:corr}
\end{figure}

The correlation values for each regime and month are illustrated in figure \ref{fig:corr}. What stands out from this figure is that cold and weak-wind events are positively correlated over land but negatively correlated over the open sea, particularly during the coldest months. Conditions over land are the most relevant for energy applications, but coastal cities might have a complicated response. There is more variation in shallow water basins, such as the North Sea, although the events are mainly positively correlated. In general, correlations are higher in Scandinavia and Northern Europe. Furthermore, in March, correlations are also negative across Southern Europe over land.

\subsection{Joint probabilities of cold and weak-wind events} \label{joint_probabilities}

The Gaussian copula framework makes it possible to generate a joint distribution from arbitrary marginal probabilities taking as an argument the covariance matrix. In this case, the BSS was computed with respect to a model with no correlations.

\begin{figure}
    \centering
    \includegraphics[scale=0.35]{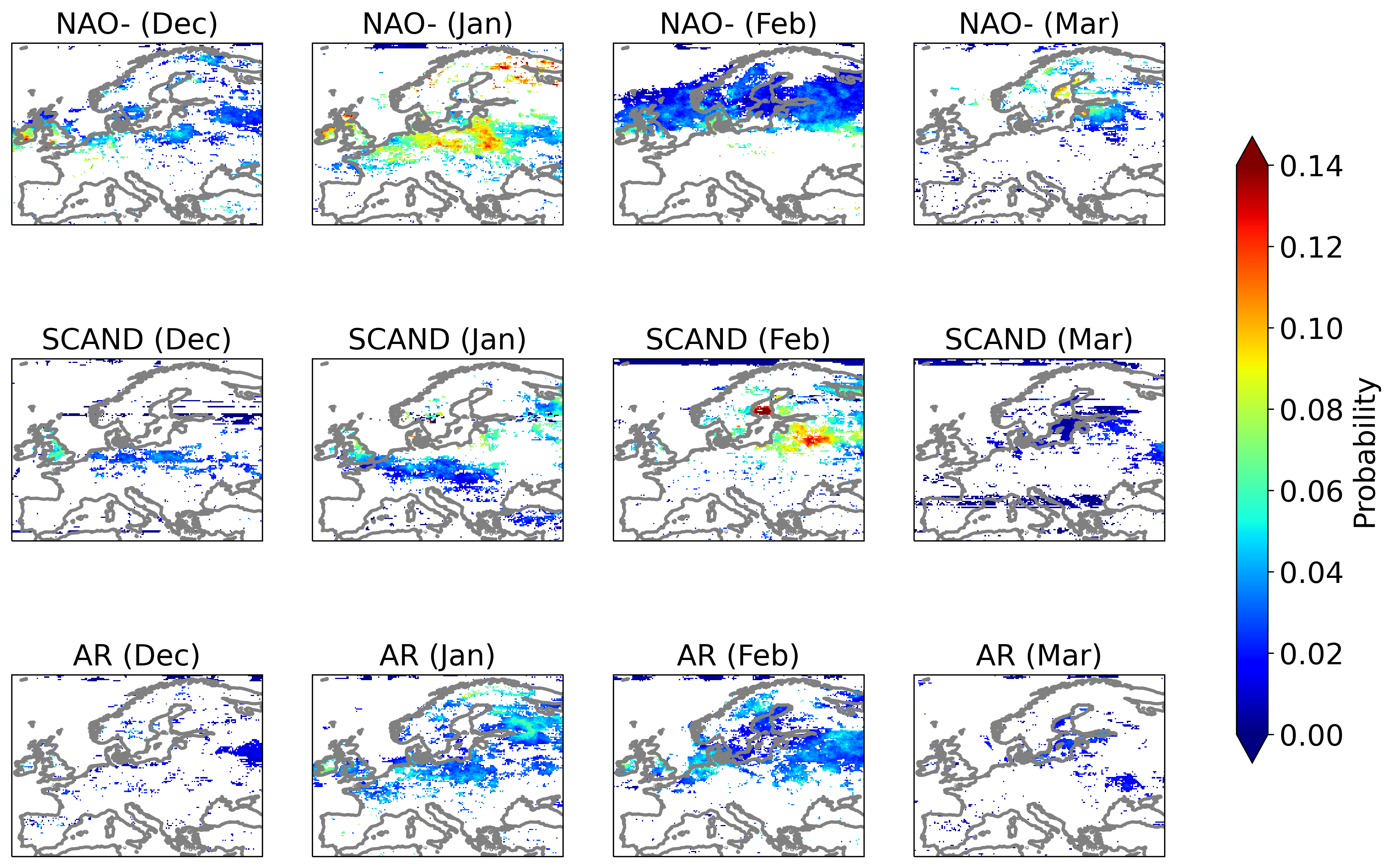}
    \caption{Joint probabilities of cold and weak-wind events where the BSS is significant. Colors show the probabilities of co-occurrence of cold and weak-wind events for each Weather Regime (WR) and month. The events were computed for daily minimum temperatures and maximum wind speeds below the $10^{th}$ percentile of the data in the extended winter season (NDJFM). Probability values where the BSS is not significant at a 10\% level are masked out. The BSS was computed with respect to a model with no correlation between cold and weak-wind events. Area of study: 35$^\circ$N--72$^\circ$N, 11$^\circ$W--40$^\circ$E. Figure based on ERA5 data (1979--2021).}
    \label{fig:copulas_joint_p_masked}
\end{figure}

Significant joint probabilities of cold and weak-wind events are shown in figure \ref{fig:copulas_joint_p_masked}. Here, we mean by significant that the correlations estimated with the rank likelihood method \cite{sbgcop} improve the results. This occurs in smaller regions of Northern and Central Europe when NAO-, SCAND, or AR conditions are dominant, in particular, in January and February. However, probabilities higher than 0.1, where the BSS is significant, occur only for NAO- in January and SCAND in February at specific locations. For NAO+ (not shown), the correlations do not significantly improve the results.  

\subsection{Estimated number of days with compound cold and weak-wind events} \label{number_days}

By considering the number of days in each month, we can translate the joint probabilities into the number of days we expect compound events.

\begin{figure}
    \centering
    \includegraphics[scale=0.35]{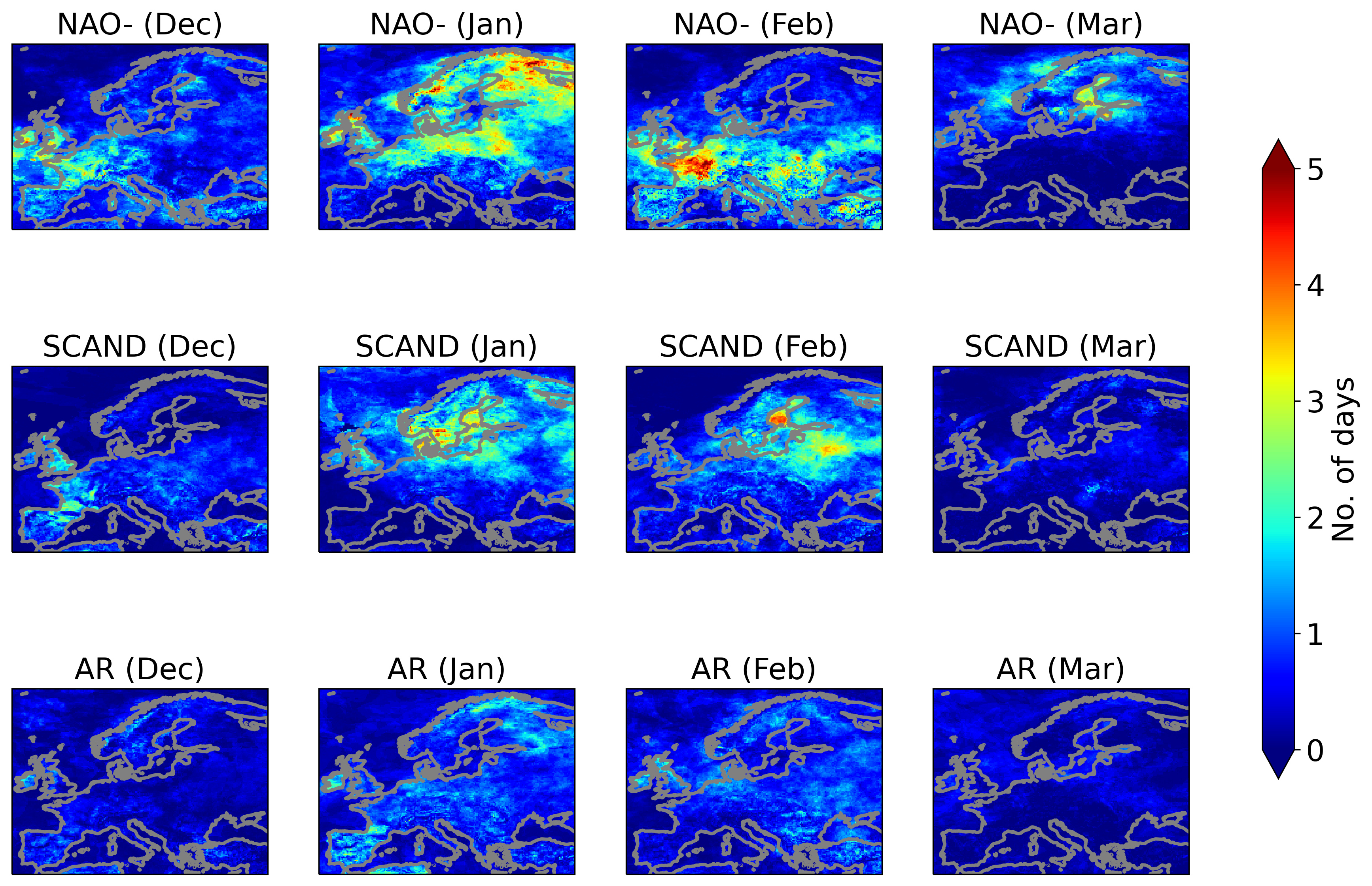}
    \caption{Number of compound cold and weak wind events. The events were computed as the daily minimum temperatures and maximum wind speeds below the $10^{th}$ percentile of the data in the extended winter season (NDJFM). Area of study: 35$^\circ$N--72$^\circ$N, 11$^\circ$W--40$^\circ$E. Figure based on ERA5 data (1979--2021).}
    \label{fig:ndays}
\end{figure}

Figure \ref{fig:ndays} shows that the number of events varies with the regime, month, and region considered. Two regimes stand out during the coldest months: NAO- and SCAND. In December, at a country level, compound events are expected to occur 0--1 on average in most countries. For NAO-, this number rises to 2 days in France, Belgium, Switzerland, and the UK and 3 days in Ireland. In January, during NAO- and SCAND, several countries in Northern Europe and Scandinavia experience compound events on average 2--3 days. We also see that these two patterns have a strong north-south gradient. Contrarily, when AR dominates, there is no evident gradient, and the number of compound events is 1--2 in the Iberian Peninsula, Ireland, the UK, and Scandinavia. The highest number of average days per country is estimated for France, Luxembourg, and Belgium during NAO- conditions in February, and local values of 5--6 days are observed in France. Also, in February, but during SCAND, an average of 2 days is expected in the Baltic countries and Poland. Finally, in March, the average number of expected compound events per country drops again to 0--1, except in Estonia, where the average is 2 days. We highlight the spatial and temporal variability of the results. Although the maximum average number of days at a country level is 3, and only for a few combinations of WR and month, figure \ref{fig:ndays} shows smaller regions where we expect compound events 5 days a month in January and February associated with blocking conditions.

Overall, figure \ref{fig:ndays} supports what previous studies have found: a) that large-scale circulation patterns influence the energy demand \cite{ely_et_al_2013, thornton_et_al_2017, Bloomfield_2018} and the wind power generation over Europe \cite{brayshaw_et_al_2011, zubiate_et_al_2016, cradden_et_al_2017, Bloomfield_2018}, and  b) that blocking patterns are associated with anomalous cold and low-wind speed conditions over Northern and Central Europe, which, in turn, increase the demand and reduce the wind power generation \cite{ely_et_al_2013, cradden_et_al_2017, Bloomfield_2018, ravestein_et_al_2018, jerez_trigo_2013, zubiate_et_al_2016}. 

\section{Discussion and conclusions} \label{discussion_and_conclusions}

Low temperatures are associated with high electricity demand during the winter, while wind speed is directly related to electricity production. The position of anomalous pressure systems and planetary waves disturb the zonal flow at 500 hPa, which, in turn, influences the progression of WRs that impact surface variables, such as temperature and wind speed. Here, we develop a Gaussian copula-based framework for modeling monthly bivariate distributions of cold weak-wind events conditioned on large-scale WRs. This flexible methodology can be adapted to different univariate distribution models.

Blocking conditions usually cause above-average demand and below-average wind and solar generation in Central and Northern Europe \cite{grams_et_al_2017, van_der_Wiel_2019}). Composites of temperature and wind speed indicate that the impacts of NAO- and SCAND on near-surface variables are associated with cold and weak-wind conditions \cite{bloomfield_etal_2020, van_der_Wiel_2019}. Thus, more extreme events are expected for these two clusters. We confirmed this hypothesis by analyzing the number of past events divided by the possible outcomes, which provides insight into how many events we can expect given the defined threshold instead of relying on the departures from average conditions, as most studies do. 

First, we estimated the correlations with the 'sbgcop' package and modeled the marginals probabilities with logistic regressions using two sets of binary predictors, one for the months in the winter season and one for the WRs. Permutation tests were used to assess the performance of the logistic models in terms of the BSS computed with respect to a seasonality model that only considers the set of monthly predictors. The results indicate that incorporating the WR information into the models improves the performance over large geographical areas. This means that large-scale systems impact the occurrence of extreme temperature and wind speed events and, consequently, the peaks of demand and production of energy. 

Then, we computed the joint probabilities with Gaussian copulas using the estimated correlations and marginal probabilities. The BSS for joint probabilities reveals an improved skill in smaller regions when accounting for the associations between the event variables, mainly in latitudes higher than 45N. It follows from the calculation of the joint probabilities that the co-occurrence of cold and weak-wind events is restricted to specific geographical areas, depending on the month and the predominant WR. These probabilities have been converted to the number of days we expect compound events for a given month and WR. Overall, the highest number of days is expected during NAO- and SCAND from December to February, in particular, in coastal and inland regions in Northern Europe with high wind farm density. The results also expose the need to consider each month separately when analyzing compound events of cold and weak winds. We conclude that the gridded results exhibit a spatial variability of extreme events that is not captured by the national aggregates presented in other studies \cite[e.g.,][]{bloomfield_et_al_2021}. 

There is an unavoidable degree of uncertainty affecting reanalysis datasets; the most significant disagreements in DJF are encountered within continental areas \cite{ramon_et_al_2019}. Representativeness can also be a problem because wind farms are often located in places where the wind is stronger than its surroundings, and thus the mean value of the grid box might be inaccurate. Moreover, one of the challenges of working with wind speed data from reanalyses is that they are available at 10 meters, whereas typical hub heights are 80-120 meters. Therefore, an extrapolation method is typically used to estimate hub height winds from surface winds. The power law \cite[e.g., ][]{van_der_wiel_2019_b} was tested for the computation of marginal weak-wind probabilities (not shown) but discarded because it adds even more uncertainty to the already biased winds. 

The scope of this study was limited to the analysis of meteorological variables associated with poor conditions for onshore wind power generation during periods of high demand. It is well known that the relationship between wind speed and wind power is nonlinear. Still, wind power is largely determined by wind speed, and our results are, therefore, useful for understanding problems of undersupply. Furthermore, we believe our work will contribute to designing more robust energy systems for periods of co-occurrence of low wind energy generation and high demand.

The skill could be improved by modeling the marginal probabilities with more complex algorithms and considering more variables. Future studies should also explore how solar and hydro can complement wind power generation in situations with high probabilities of co-occurrence of cold and weak-wind events. We also recommend examining the applicability of this methodology to energy demand and energy shortfall variables commonly used in the literature \cite[e.g., ][]{van_der_wiel_2019_b, bloomfield_etal_2020, otero_et_al_2021}. These variables have been studied in a copula-based framework \cite{otero_et_al_2021}, but with a focus on seasonal values aggregated at a country level. The pronounced subseasonal and spatial variability observed in the present study suggests that decision-makers would further benefit from monthly results on a grid when analyzing energy demand and shortfall variables.

In summary, the results confirm that the dependence between the wind speed and temperature event variables plays an essential role in modeling compound meteorological events and that low-frequency circulation patterns control parts of the distribution of extreme events. We also provide evidence of strong intraseasonal and spatial variability of compound events. 

\section{Acknowledgments}\label{acknowledgments}

The authors have received funding from the European Union's Horizon 2020 research and innovation program under grant agreement No. 776787 (S2S4E). This work has also been supported by the Research Council of Norway through grants 303411 (Machine Ocean) and 309562 (Climate Futures).

This research was conducted using computational resources provided by the Norwegian Meteorological Institute and the University of Oslo. 
 
The authors would also like to thank Frode Stordal for interesting discussions and Jean Rabault for encouraging the publication of this research.

\bibliographystyle{unsrtnat}
\bibliography{references}  


\appendix
\section{Composites}\label{sec:appendix_composites}

\begin{figure}[htp!]
    \centering
    \includegraphics[scale=0.3]{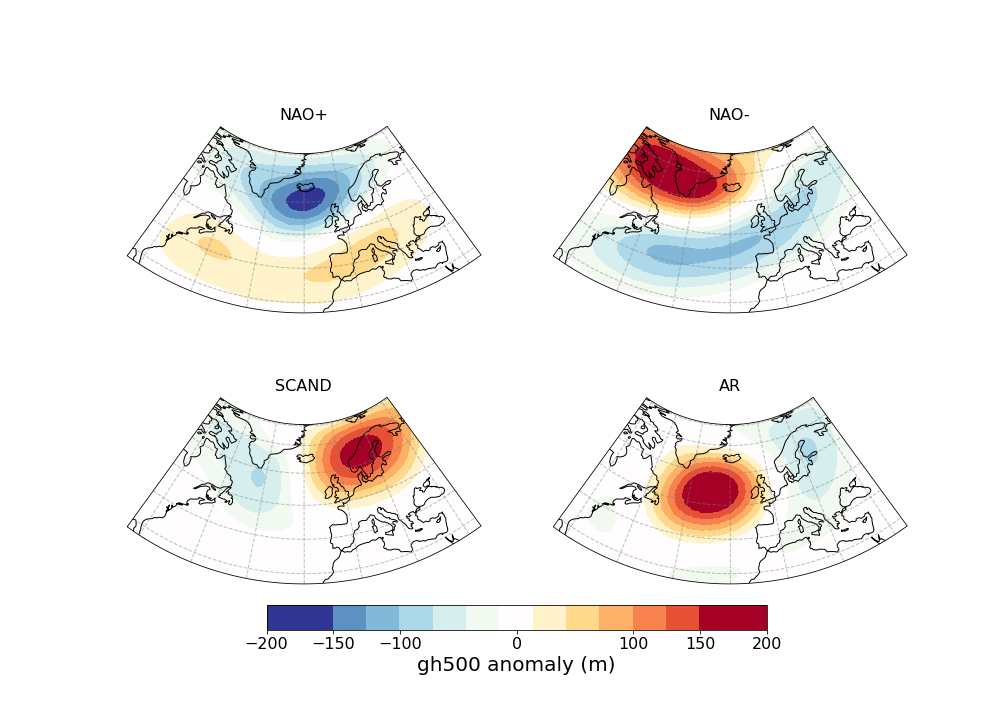}
    \caption{Four winter regimes of atmospheric circulation in the North Atlantic-European domain obtained by clustering geopotential height data at 500 hPa. The four patterns obtained resemble the traditional positive phase of the North Atlantic Oscillation (NAO+), the negative phase of the North Atlantic Oscillation (NAO-), the Scandinavian Blocking (SCAND), and the Atlantic Ridge (AR) patterns. Colors show the Z500 anomaly in meters. Area of study: 20$\circ$N--80$\circ$N, 90$\circ$W--$60\circ$E. Figure based on ERA5 data (NDJFM, 1979--2021).}
    \label{fig:WRs}
\end{figure}

The Z500 anomaly fields for each of the four WRs are illustrated in figure \ref{fig:WRs}. Although not identical, all four patterns resemble the traditional patterns computed by the NOAA: NAO+, NAO-, SCAND, and AR. For this reason, we will refer to the WRs obtained with clustering techniques with the same names.

For the sake of comparison, composites of temperature and wind speed anomalies were also computed for each WR, replicating the results obtained by other studies \cite{bloomfield_etal_2020, van_der_Wiel_2019} and showing the robustness of the method. Anomalous low surface temperatures over Europe are common during blocking conditions, in this context, during NAO- and SCAND.

\begin{figure}
    \centering
    \includegraphics[scale=0.3]{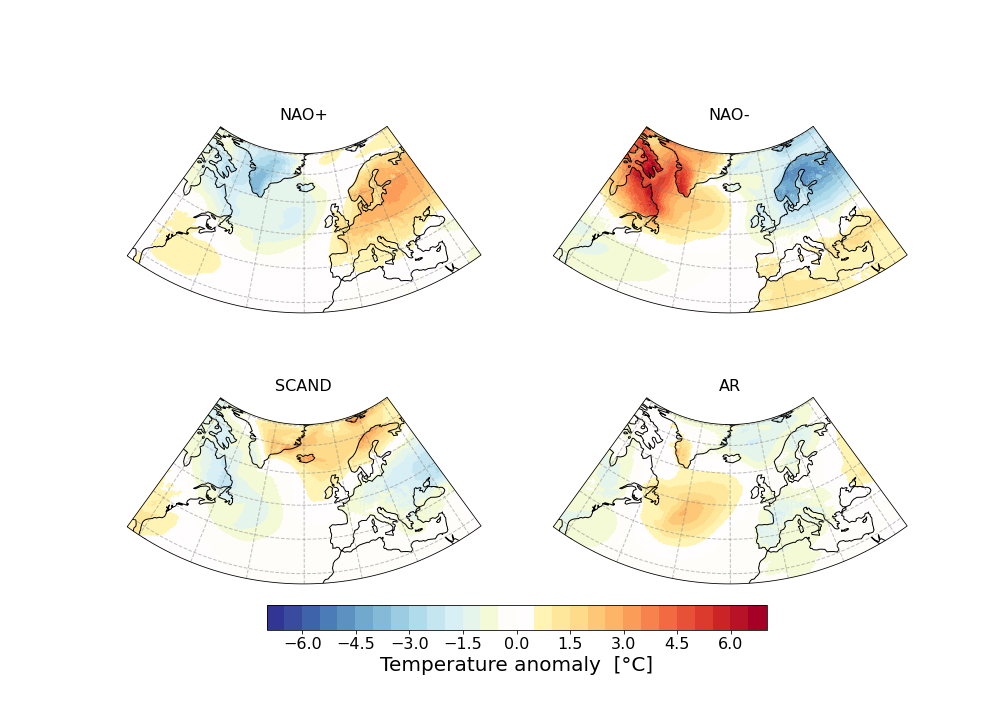}
    \caption{Mean 2m temperature impacts of the four weather regimes (WRs). Colors show temperature anomalies ($^\circ C$). Area of study: 27$^\circ$N--81$^\circ$N, 85.5$^\circ$W--45$^\circ$E. Figure based on ERA5 data (NDJFM, 1979--2021).}
    \label{fig:composites_t}
\end{figure}

As expected, the regimes with the strongest impact on surface weather are NAO- and NAO+. The response of these regimes in both the temperature and the wind speed fields is nearly symmetric. The impacts of the NAO on surface temperature (figure \ref{fig:composites_t}) consists of a zonal dipole with centers over Northern Europe and the Labrador Sea. During NAO-, temperature anomalies are, on average, cold across Northern Europe, Scandinavia, and Northwestern Russia, but warm across the Mediterranean countries. This results from above-average geopotential heights over Iceland, allowing cold air to drain from high latitudes, and a lower than normal pressure system over the Azores high. These combined effects decrease the pressure gradient across the North Atlantic. The opposite conditions yield for NAO+, meaning that warm anomalies cover the entire continent. 

A north-south gradient of mean temperature anomalies can be observed during SCAND conditions in figure \ref{fig:composites_t}. It is, however, less prominent and has the opposite sign of the gradient associated with NAO-. The anticyclonic circulation associated with the SCAND imposes warm conditions over Scandinavia and Britain, extending over the Norwegian Sea to Greenland, as a cold polar continental air mass develops across the rest of Europe. On the other hand, the lowest climatological departures of minimum daily temperatures are observed during AR days. This pattern consists of two centers of warm anomalies located over the North Atlantic and Northwestern Russia, and cold anomalies over the European continent spanning the Norwegian Sea and Greenland.

\begin{figure}
    \centering
    \includegraphics[scale=0.3]{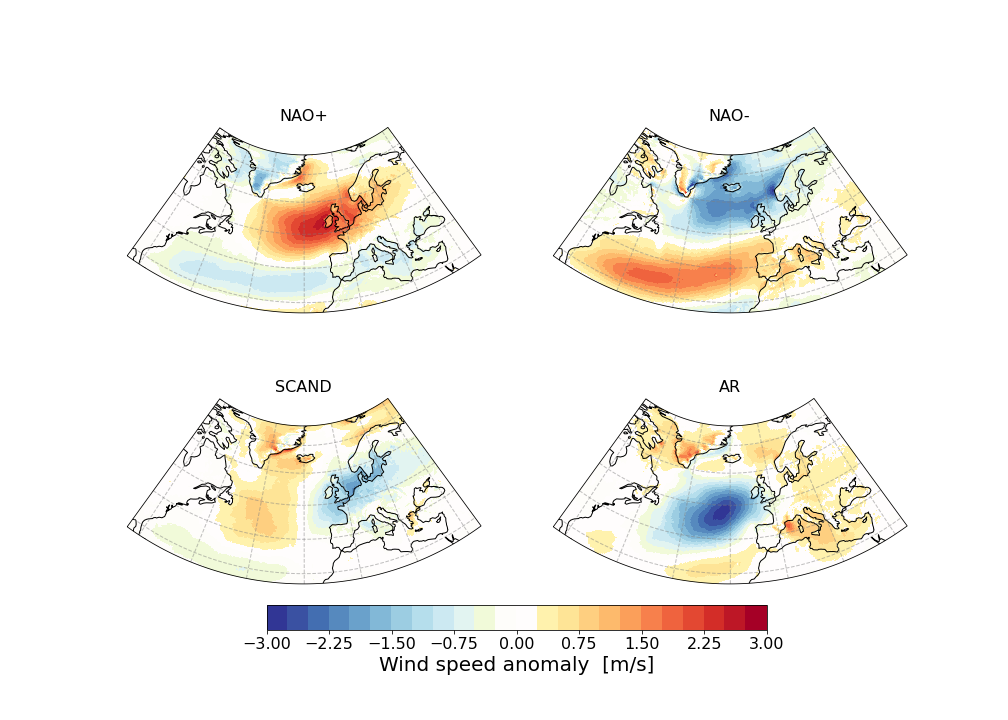}
    \caption{Mean 10m wind speed impacts of the four weather regimes (WRs). Colors show wind speed anomalies ($m/s$). Area of study: 27$^\circ$N--81$^\circ$N, 85.5$^\circ$W--45$^\circ$E. Figure based on ERA5 data (NDJFM, 1979--2021).}
    \label{fig:composites_ws}
\end{figure}

Figure \ref{fig:composites_ws} shows the typical surface imprint of the four WRs on wind speed anomalies. Weak winds are expected over Europe during blocking conditions (e.g., NAO- and SCAND),  when a high pressure system typically dominates the circulation over Europe. During NAO, a meridional wave train extends across the Atlantic and Western Europe with an opposite sign for the positive and negative phases. During NAO+, countries in Northern Europe are typically affected by high anomalous wind speeds, whereas Mediterranean countries experience anomalous low wind conditions; the opposite yields for the NAO-. The SCAND is characterized by anticyclonic anomalies resulting in anomalous weak winds centered over Britain and affecting most European countries. Similarly, the impact of the AR on the wind speed anomaly field consists of a monopole of weak winds but, in this case, with a center located in the North Atlantic associated with strong wind anomalies spanning the European continent.


\section{Marginal probabilities and significant skill scores}\label{sec:appendix_unmasked_probabilities}
To compute the joint probabilities of compound events, we need to compute the marginal probabilities in addition to the correlations.

\begin{figure}
    \centering
    \includegraphics[scale=0.35]{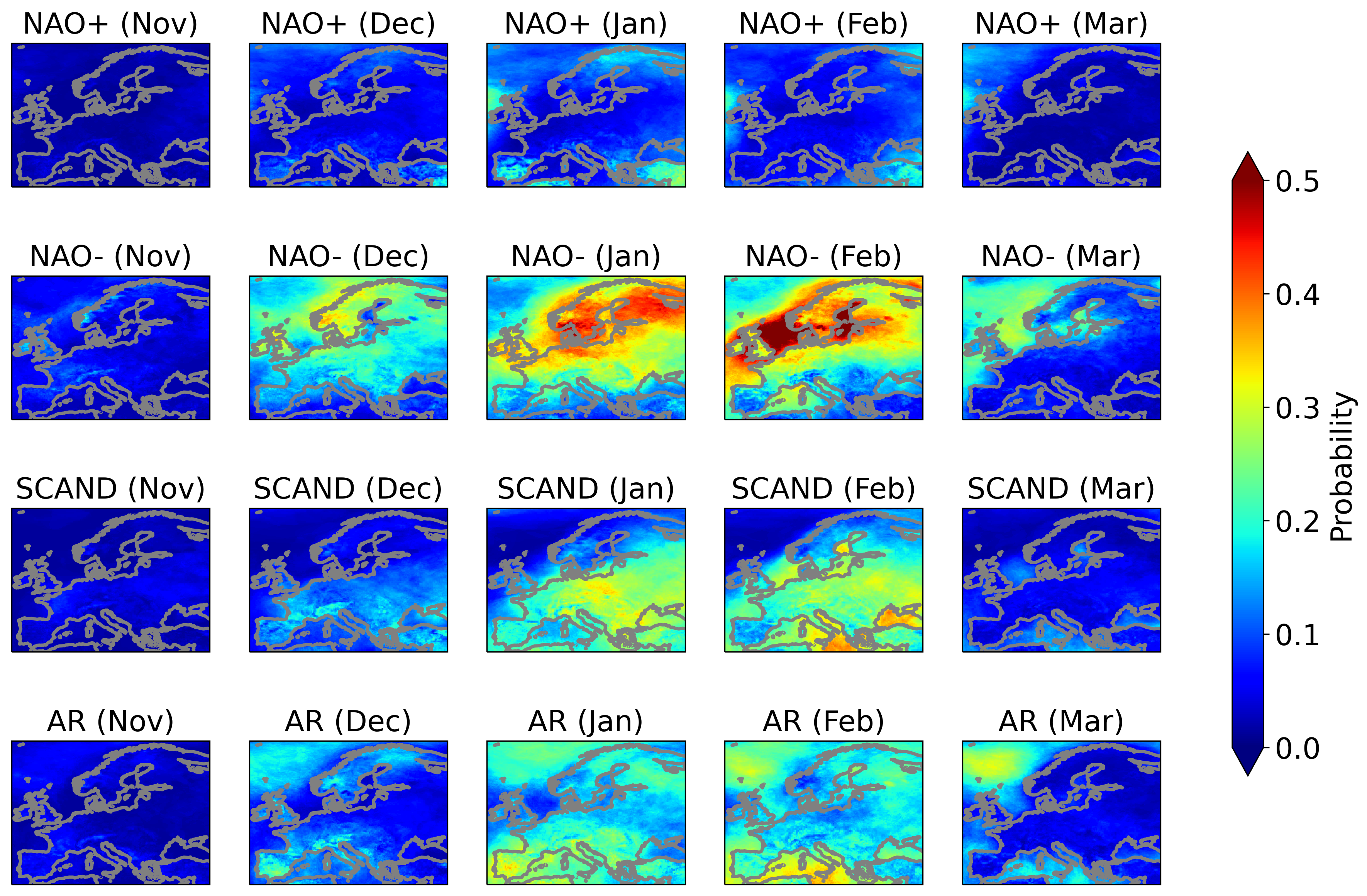}
    \caption{Marginal probabilities of cold events.
     Colors show the probabilities of cold events for each Weather Regime (WR) and month computed with a logistic regression model. The events were defined as daily minimum temperatures below the $10^{th}$ percentile computed for the entire winter season (NDJFM). Area of study: 35$^\circ$N--72$^\circ$N, 11$^\circ$W--40$^\circ$E. Figure based on ERA5 data (1979--2021).}
    \label{fig:logistic_t2m}
\end{figure}

Figure \ref{fig:logistic_t2m} indicates that temperatures lower than the $10^{th}$ percentile are rarely observed in March. Probabilities are, in general, higher over land and coastal regions than over the open ocean in December and January. The opposite occurs at the end of the season, in February and March. The logistic regressions estimate that blocking conditions lead to cold events more than 20\% of the days in a given regime and month over extensive regions during the coldest months. Low temperatures can still be expected over 30\% of the days in March under NAO- and AR conditions, which we speculate is due to a slow response of the oceans. 

The NAO- is the WR with the strongest temperature signal, followed by the SCAND and the AR. In contrast, the NAO+ shows the lowest probabilities of low temperatures. Days classified as NAO- show an increase in the number of cold events from December to February, reaching values as high as  $p_{X_{(Feb, NAO-)}} = 0.58$, and a spatial average equal to $p_{\mu_{(Feb, NAO-)}} = 0.27$. Values are particularly high over the North Sea, the Baltic Sea, and the Bay of Biscay, and we can expect cold events more than 40 \% of the days in northern latitudes in January and February.

An important difference between the NAO- and the SCAND, besides the magnitude, is the spatial distribution. During NAO-, high latitudes are affected by high probabilities, while low latitudes are affected by low probabilities. The opposite takes place during SCAND. Furthermore, when SCAND is dominant, large continental areas are expected to experience low temperatures in January and February. For this cluster, the first cold events of the season appear in December in lower latitudes over land. The highest probabilities over land are observed in January, with a maximum equal to $p_{X_{(Jan, NAO-)}} = 0.36$ in Eastern Europe, but over the sea, the maximum is reached in February and equals $p_{X_{(Jan, NAO-)}} = 0.38$.

The AR is characterized by a center of relatively high probabilities situated over the Norwegian Sea, which intensifies from December to March, and high probabilities in the Iberian Peninsula and the Mediterranean Sea. Moreover, a zonal band of low probabilities extends approximately from 45N to 60N.

\begin{figure}
    \centering
    \includegraphics[scale=0.35]{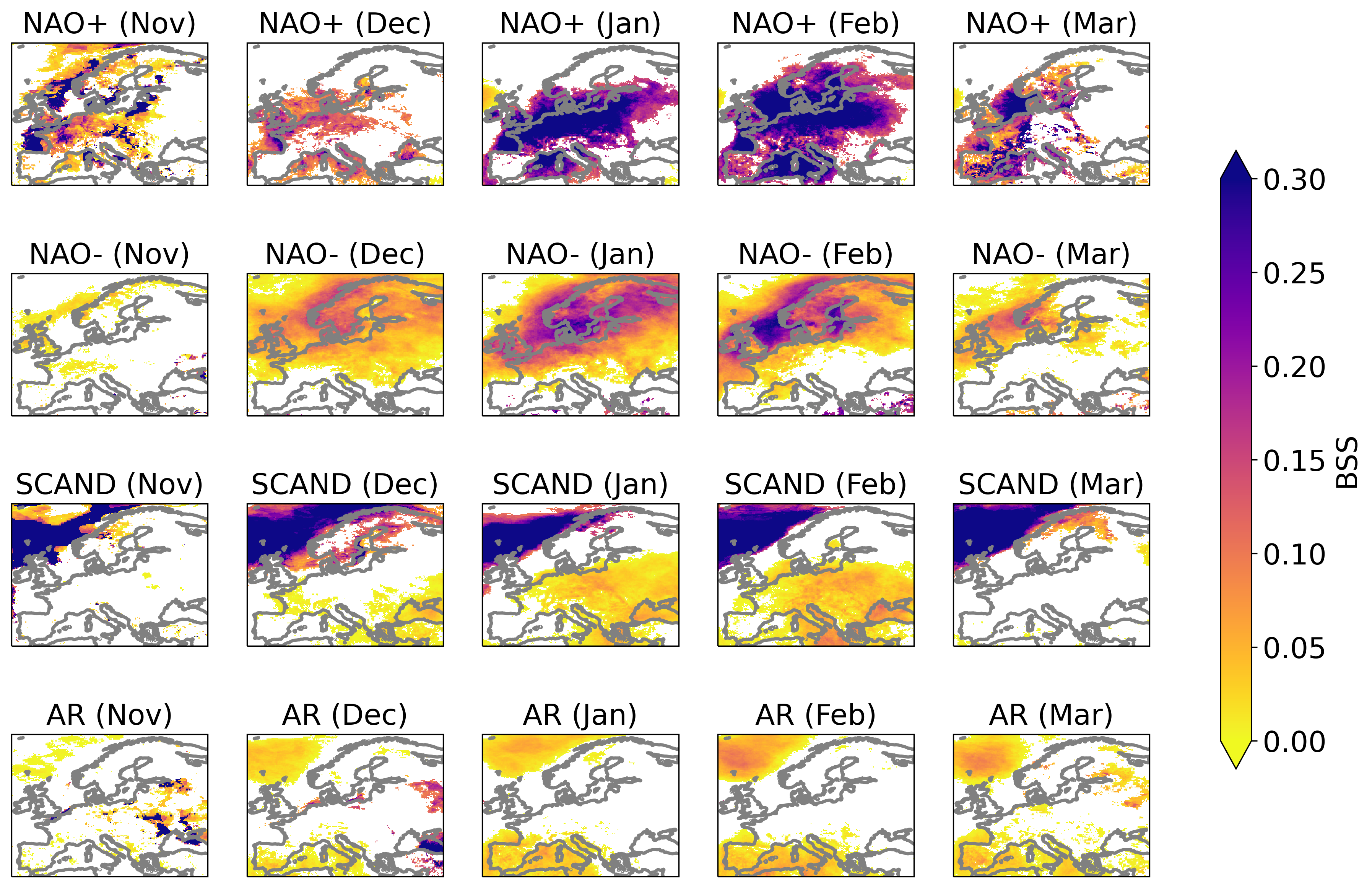}
    \caption{Significant BSS computed for probabilities of cold events for each Weather Regime (WR) and month. The marginal probabilities of cold events were estimated with a logistic regression model. The BSS was computed with respect to a climatology model. Colors show positive significant BSS values at the 10\% level. Area of study: 35$^\circ$N--72$^\circ$N, 11$^\circ$W--40$^\circ$E. Figure based on ERA5 data (1979--2081).}
    \label{fig:bss_t2m_masked}
\end{figure}

Figure \ref{fig:bss_t2m_masked} shows the BSS computed with respect to a seasonality model that only considers the set of monthly predictors. Positive and significant values indicate that WRs impact extreme temperatures and, consequently, the peaks of demand. In other words, providing the WRs as input to the logistic regression models leads to a significant improvement compared to only accounting for the months. Furthermore, figure \ref{fig:bss_t2m_masked} shows that NAO+ is the regime with the highest BSS over land. On the other hand, SCAND scores highest over the North Atlantic, reaching values higher than 0.9 every month of the season. Notice that these regions are characterized by very low probabilities of cold events (see figure \ref{fig:logistic_t2m}). Moreover, the highest BSS during SCAND over the land occurs during the coldest months, in regions where high probabilities of low temperatures are assigned. Meanwhile, NAO- scores highest during the coldest months in Scandinavia, Britain, and the North Sea, where we expect the more cold events. Finally, the AR shows a significant improvement in lower latitudes and the North Atlantic. In general terms, the regions affected by high probabilities of occurrence of cold events (see figure \ref{fig:logistic_t2m}) have a positive and significant BSS. 
 
\begin{figure}
    \centering
    \includegraphics[scale=0.35]{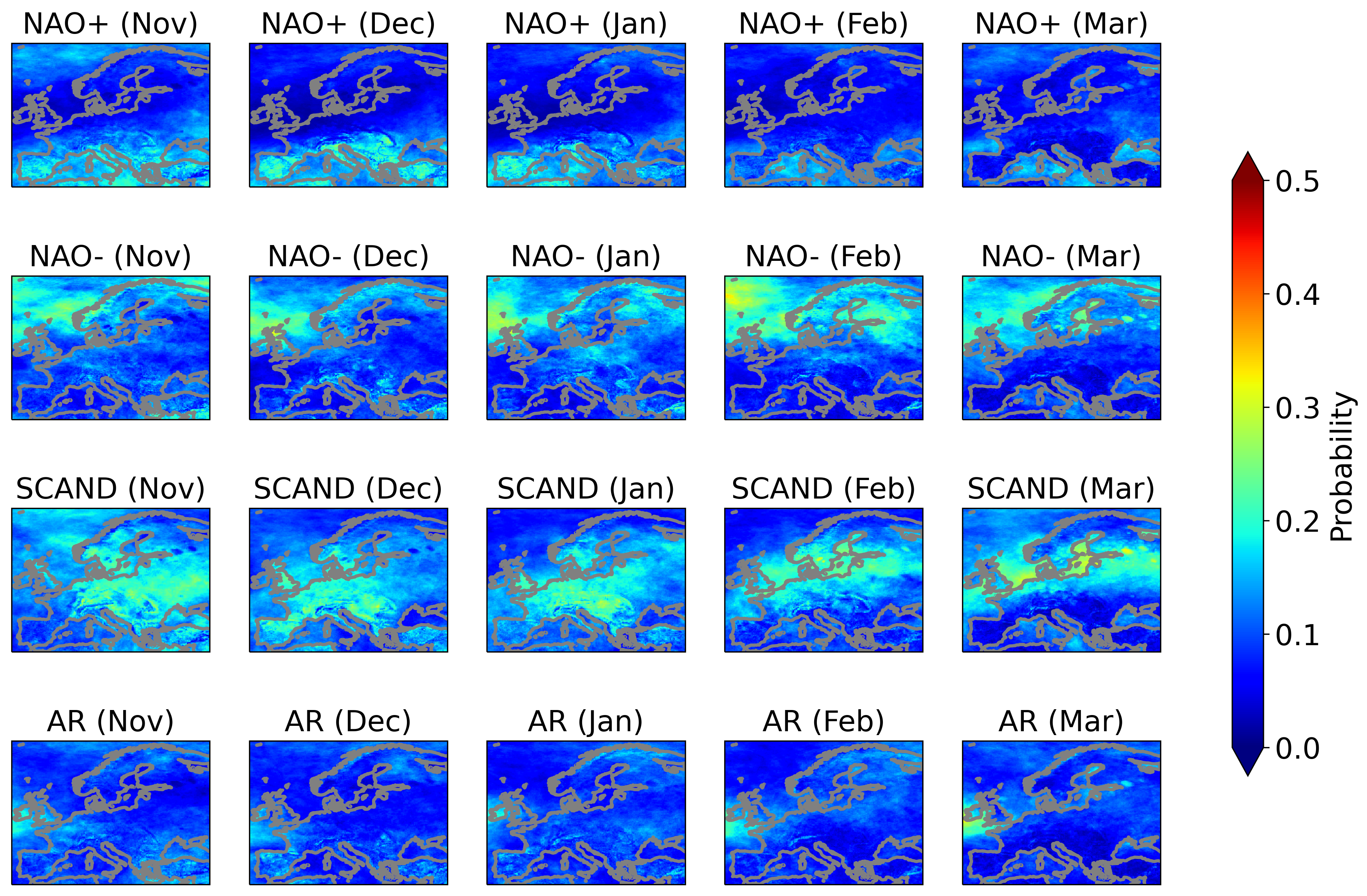}
    \caption{Marginal probabilities of weak-wind events.
     Colors show the probabilities of weak-wind events for each Weather Regime (WR) and month computed with a logistic regression model. The events were defined as daily maximum wind speeds below the $10^{th}$ percentile computed for the entire winter season (NDJFM). Area of study: 35$^\circ$N--72$^\circ$N, 11$^\circ$W--40$^\circ$E. Figure based on ERA5 data (1979--2021).}
    \label{fig:logistic_ws10m}
\end{figure}

Probabilities of low maximum daily wind speed are shown for each month and WR in figure \ref{fig:logistic_ws10m}. These are mainly dominated by the zonal flow, and have a small intraseasonal and spatial variability compared to cold events. The signal is also noisier, responding to changes in the terrain. Probabilities are predominantly higher for NAO- and SCAND compared to AR and NAO+. According to the logistic model, both NAO- and SCAND exhibit, on average, weak-wind probabilities between 0.1 and 0.15. However, we can identify regions where weak-wind events are expected 20\% of the days.

The highest probabilities over land are observed in Central and Northern Europe during the SCAND from November to January. Contrarily, over the ocean, the logistic model predicts more days with calm winds in March. The Arctic and Southern Europe are affected by low probabilities during SCAND. Furthermore, all the moths show local maximum values above 0.3. On the other hand, the NAO- exhibits a zonal pattern with lower probabilities in Central and Southern Europe, and higher probabilities in Northern Europe. Seasonal changes during NAO- are small, although probabilities in the North Atlantic increase from November to February, and decrease in March. The NAO+ has a weaker and opposite pattern, with higher probabilities in the south and lower probabilities in the north. This pattern is observed from November to February. Meanwhile, during AR conditions, a persistent maximum is located over Britain throughout the extended winter season. 

\begin{figure}
    \centering
    \includegraphics[scale=0.35]{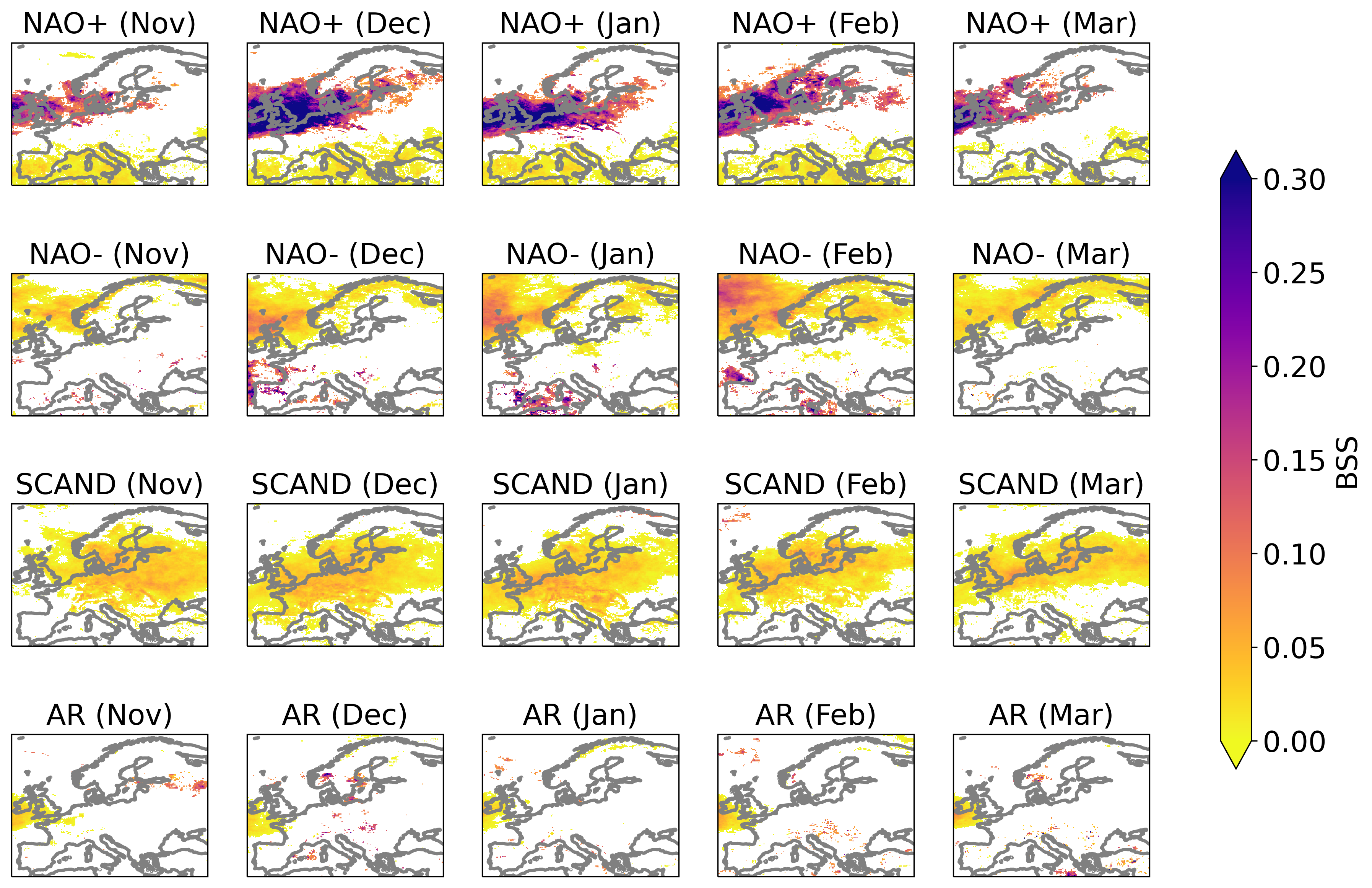}
    \caption{Significant BSS computed for probabilities of weak-wind events for each Weather Regime (WR) and month. The marginal probabilities of weak-wind events were estimated with a logistic regression model. The BSS was computed with respect to a climatology model. Colors show positive significant BSS values at the 10\% level. Area of study: 35$^\circ$N--72$^\circ$N, 11$^\circ$W--40$^\circ$E. Figure based on ERA5 data (1979--2081).}
    \label{fig:bss_ws10m_masked}
\end{figure}

As for the marginal wind speed probabilities, the highest significant BSS values are associated with NAO+ in Northwestern Europe (see figure \ref{fig:bss_ws10m_masked}). Still, the logistic model that considers large-scale circulation information outperforms the seasonality model in the regions with the highest weak-wind probabilities associated with blocking patterns. In particular, during NAO-, BSS are positive across the Mediterranean and North Atlantic. The maximum score is $BSS_{X_{(Jan, NAO-)}} = 0.99$. Meanwhile, during SCAND, the significant values are located over Northern and Central Europe. AR conditions lead to no significant BSS pattern. 

To synthesize, cold events and low wind speed events are associated with the blocking patterns NAO- and SCAND. While temperature events are restricted to the coldest months, wind speed events show lower intraseasonal variations. Positive and significant scores reveal that large-scale atmospheric circulation affects the occurrence of cold and weak-wind events, i.e., the WRs are important predictors of marginal probabilities. Overall, we observe that areas with high probabilities also have significant BSS. Given that November is the warmest month and some regions do not register any cold event, we show only the results from December to March in the following.

\section{Joint probabilities and significant skill scores}\label{sec:appendix_unmasked_joint_probabilities}
Provided the correlations and the marginal probabilities shown above, we can compute the joint probabilities of compound events with Gaussian copulas.

\begin{figure}
    \centering
    \includegraphics[scale=0.35]{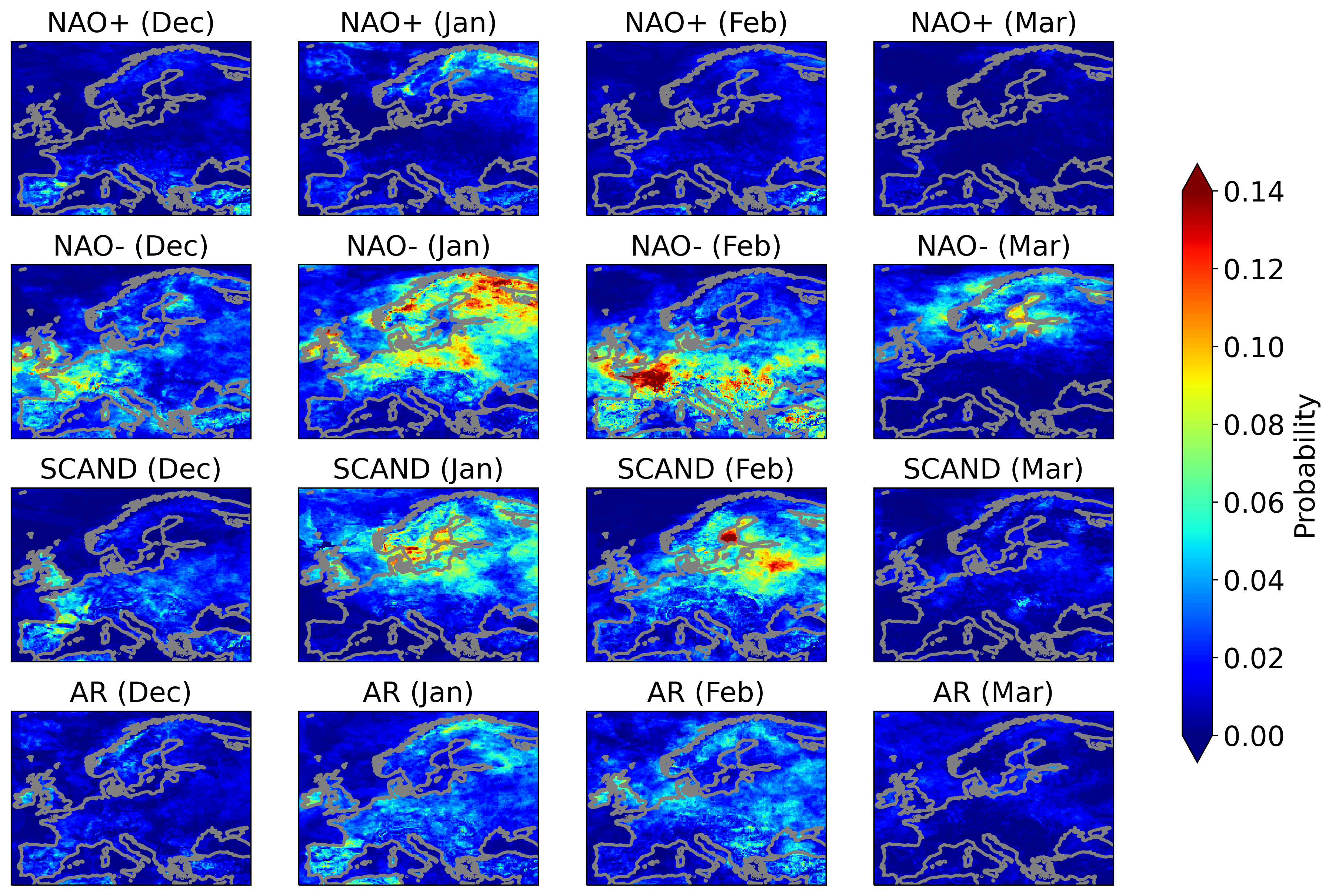}
    \caption{Joint probabilities of compound cold and weak-wind events for each weather regime (WR) and month. Colors show the probabilities of co-occurrence of cold events and low wind speed events computed with the copula model. The events were computed as the daily minimum temperatures and maximum wind speeds below the $10^{th}$ percentile of the data in the extended winter season (NDJFM). Area of study: 35$^\circ$N--72$^\circ$N, 11$^\circ$W--40$^\circ$E. Figure based on ERA5 data (1979--2081).}
    \label{fig:copulas_joint_p}
\end{figure}

Figure \ref{fig:copulas_joint_p} shows the joint probabilities for each WR. It is important to highlight the strong intraseasonal variability of compound events. The strongest signal is registered during NAO- conditions, from December to March, and during SCAND from December to February, with values above 0.1. A much weaker signal is observed when AR is dominant in January and February. In December, both NAO- and SCAND exhibit higher joint probabilities in Western Europe. On the other hand, in January, both regimes show probabilities higher than 0.5 across a large region covering Northern Europe and Scandinavia. Only days classified as NAO- present joint probabilities higher than 0.5 in March.

\begin{figure}
    \centering
    \includegraphics[scale=0.35]{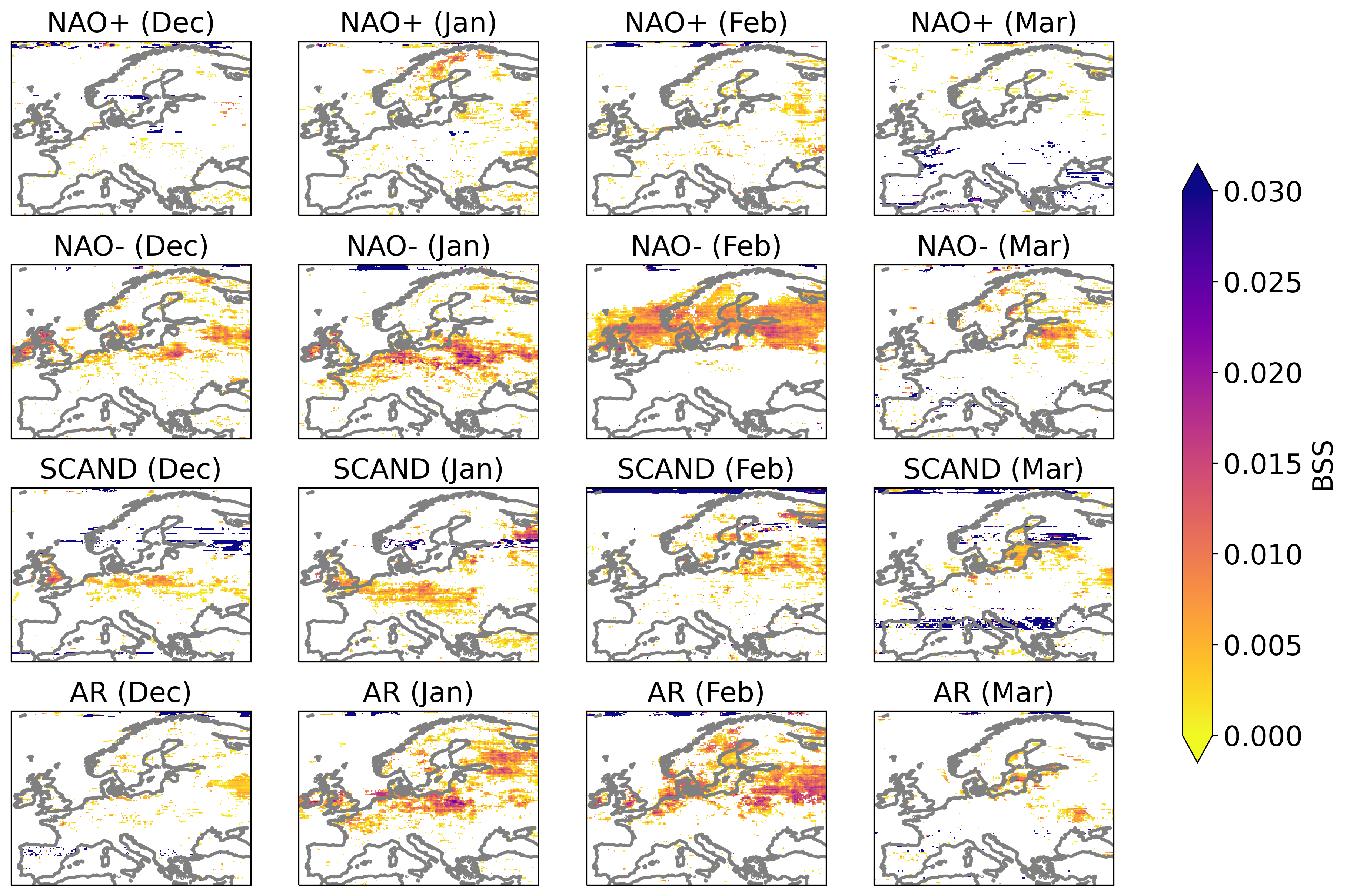}
    \caption{Positive significant BSS at the 10\% level for joint probabilities modeled with Gaussian copulas and logistic regressions, aggregated by Weather Regime (WR) and month. An independent model, with no correlation between the temperature and the wind speed events, was employed as the reference model. Colors show significant BSS. Area of study: 35$^\circ$N--$72^\circ$N, 11$^\circ$W--40$^\circ$E. Figure based on ERA5 data (NDJFM, 1979--2021).}
    \label{fig:bss_joint_copulas_pos}
\end{figure}

Significant and positive BSS values of joint probabilities are illustrated in figure \ref{fig:bss_joint_copulas_pos}. In this case, the BSS was computed with respect to a Gaussian model with correlations equal to zero. The transition months and the NAO+ regimes are excluded because they register very few cold events. Figure \ref{fig:bss_joint_copulas_pos} shows that including the correlations in the Gaussian copula framework improves the prediction skill in Northern Europe and Scandinavia. The fact that there are extended areas with significant positive scores proves the potential of the Gaussian copula framework in estimating joint meteorological events and the importance of considering associations between variables when modeling joint distributions. Notice that the joint probabilities are very low in Southern Europe for most of the WRs, where there is no improvement in terms of the skill score.

\end{document}